\documentclass[conference]{IEEEtran}
\IEEEoverridecommandlockouts

\usepackage{cite}
\usepackage{amsmath,amssymb,amsfonts}
\usepackage{algorithmic}
\usepackage{graphicx}
\usepackage{textcomp}
\usepackage{xcolor}
\usepackage{multirow}
\usepackage{graphicx}

\def\BibTeX{{\rm B\kern-.05em{\sc i\kern-.025em b}\kern-.08em
    T\kern-.1667em\lower.7ex\hbox{E}\kern-.125emX}}
\begin{document}

\title{TSPE: Task-Specific Prompt Ensemble for Improved Zero-Shot Audio Classification\\}

\author{\IEEEauthorblockN{Nishit Anand$^{1}$ \quad Ashish Seth$^{1}$ \quad Ramani Duraiswami$^{1}$ \quad Dinesh Manocha$^{1}$}
\IEEEauthorblockA{$^{1}$University of Maryland, College Park, MD, USA \\
\{nishit, aseth125, ramanid, dmanocha\}@umd.edu}
}

\maketitle

\begin{abstract}
Audio-Language models (ALMs) excel in zero-shot audio classification, a task where models classify previously unseen audio clips at test time by leveraging descriptive natural language prompts. We introduce \emph{TSPE} (Task-Specific Prompt Ensemble), a simple, training-free hard prompting method that boosts ALMs' zero-shot performance by customizing prompts for diverse audio classification tasks. Rather than using generic template-based prompts like ``Sound of a car'' we generate context-rich prompts, such as ``Sound of a car coming from a tunnel''. Specifically, we leverage label information to identify suitable sound attributes, such as ``loud'' and ``feeble'', and appropriate sound sources, such as ``tunnel'' and ``street'' and incorporate this information into the prompts used by Audio-Language Models (ALMs) for audio classification. Further, to enhance audio-text alignment, we perform prompt ensemble across TSPE-generated task-specific prompts. When evaluated on 12 diverse audio classification datasets, TSPE improves performance across ALMs by showing an absolute improvement of 1.23-16.36\% over vanilla zero-shot evaluation. 
\end{abstract}

\section{Introduction}

Recent progress in multimodal language models (MLMs) has greatly advanced performance across multiple modalities and tasks~\cite{b27, b28, b29, b30, b31}. Trained on large datasets of audio-caption pairs, these models gain a broad understanding of audio concepts, allowing them to classify new audio categories in a zero-shot setting. This adaptability makes Audio-Language models (ALMs) well-suited for dynamic environments with diverse and unfamiliar sounds.

Contrastive Learning-based Audio Language Models (ALMs) like Contrastive Language Audio Pre-training (CLAP)~\cite{b19} learn a shared representation space between audio and text. This helps them generalize well and have good downstream performance in tasks like audio classification. Various Audio-Language Encoders (ALEs) have been published in the literature over the past few years which perform well on tasks like audio classification and text-to-audio retrieval.

These models are pre-trained on large audio-text datasets, enabling them to generalize well in diverse audio environments. While they perform strongly in zero-shot audio classification, this success often depends on extensive audio-text pairs or additional fine-tuning. Few efforts have focused on enhancing zero-shot classification for CLAP-like models without extra training. Some approaches, like Audio Prompt Learner~\cite{b25} and TreffAdapter~\cite{b26}, improve performance but require additional training and introduce learnable parameters, which increase time and computational costs. Moreover, although effective on in-distribution tasks, they tend to perform poorly on out-of-distribution (OOD) audio classification tasks. Additionally, a common limitation in current methods is their dependence on generic prompts such as ``sound of a $<$label$>$''. We find that these prompts do not transfer effectively across different downstream tasks and often need adaptation to be meaningful. For instance, in a musical genre classification task like GTZAN~\cite{b6}, the prompt ``sound of a rock'' is unclear and does not convey the intended category, whereas modifying it to ``sound of rock \emph{music}'' provides clarity for the model to understand genres in proper context.

\textbf{Main Contribution} - 
To address this, we introduce TSPE (Task-Specific Prompt Ensemble), a training-free approach which improves the zero-shot audio classification performance of ALMs. It uses downstream task and label information to automatically generate task-specific prompts for each class label. This is important in order to capture the semantic nuances in the audio-text alignment. Then, instead of using a single vanilla prompt, it uses prompt ensembling to learn a more semantically rich representation of the prompt, which helps it to better understand the correlation between the prompt's rich textual representation and the audio representation, thus improving the performance of the ALM on downstream audio classification. Our method does not require any fine-tuning or extra training for this performance improvement, and the highlight is that it performs well on audio classification on out-of-distribution (OOD) datasets as well. We conduct extensive experiments on 12 diverse audio classification datasets and show an absolute improvement of 1.23-16.36\% over vanilla zero-shot evaluation.

\begin{figure*}[h!]
    \centering
    \includegraphics[width=1.0\textwidth]{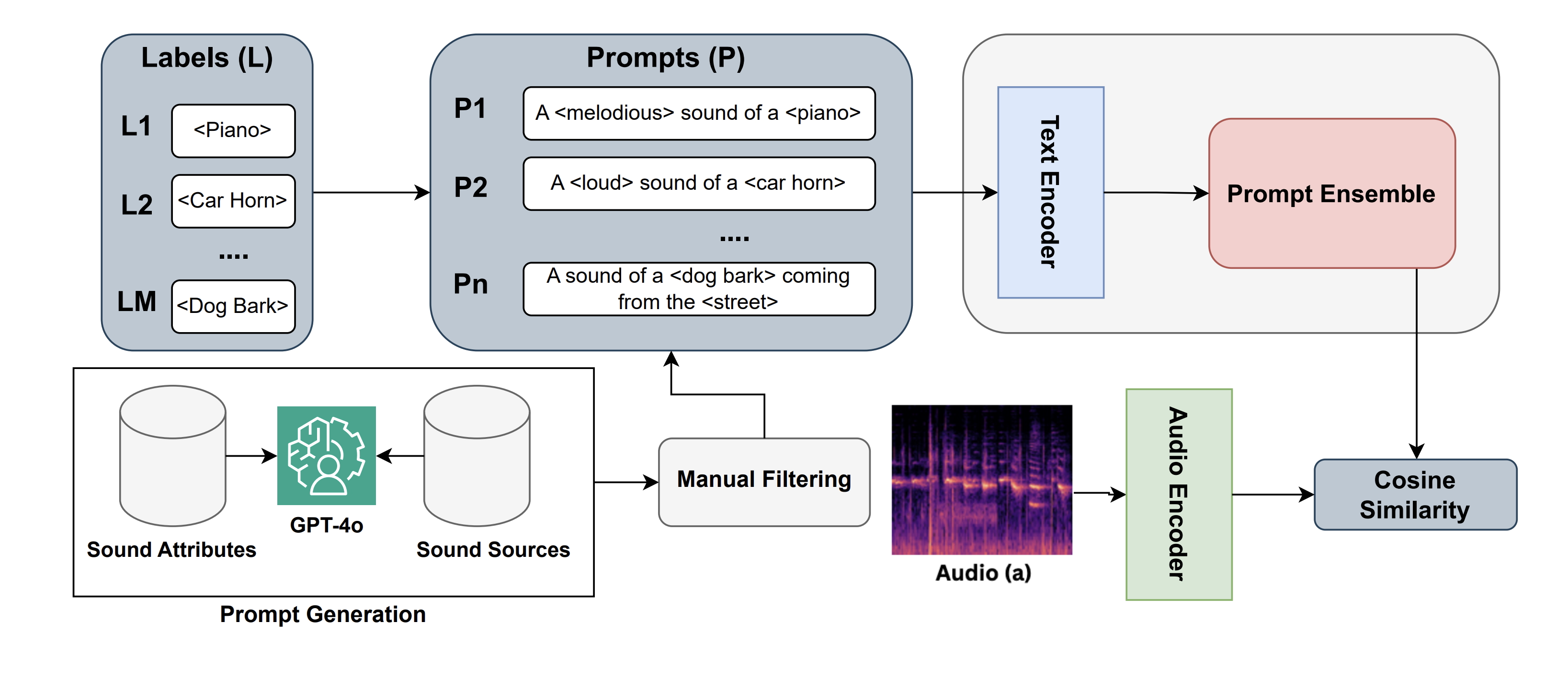}
    \caption{Illustration of TSPE Workflow: We start with a pool of sound attributes and sources to customize a set of prompt templates using GPT-4. Then, we manually select prompts relevant to the specific downstream task and pass them to the text encoder to generate representations. These representations are averaged using a prompt ensemble. Finally, we compute the cosine similarity between the audio representation and the averaged text representation.}
    \label{fig:main}
\end{figure*}

\section{Related Work}
Multimodal encoders for learning shared representations across modalities have shown significant promise. Building on contrastive pre-training methods from vision-language models like CLIP~\cite{b16}, Audio-Language Models (ALMs) have achieved state-of-the-art zero-shot performance in audio classification. Early models such as Wav2Clip~\cite{b17} and AudioCLIP~\cite{b18} focused on aligning audio representations with categorical labels, while recent approaches like CLAP~\cite{b19} map audio directly to textual descriptions, yielding substantial zero-shot gains. While hard prompting \cite{b24} has enhanced zero-shot abilities in vision-language models by using contextually rich prompt engineering~\cite{b20}, this technique remains largely unexplored for ALMs. Instead, prior ALMs have relied on compute-intensive methods, such as improving alignment objectives~\cite{b21} or scaling parameters and datasets~\cite{b19}.

\section{Methodology}
\subsection{Grouping Diverse Audio Labels}
We observe that the current prompts used for audio classification tasks often fail to capture the diversity of audio labels. For instance, generic prompts like ``sound of a $<$label$>$'' lack semantic coherence when $<$label$>$ is a musical genre such as ``rock'' or a location like ``beach''. This highlights the need to understand task-specific labels to create more effective prompts. To address this, we develop TSPE that uses the task and label knowledge to generate task-specific hard prompts. We achieve this by first categorizing labels into distinct groups based on their classification characteristics, such as the type of sound. Details of this classification are outlined below:
\begin{itemize}
    \item \textbf{Musical Instruments Recognition} This category comprises of sounds from various musical instruments played in diverse settings, such as opera, street performances, and theater. It includes a wide range of instruments, including the piano, guitar, cymbals, drums, etc. Relevant datasets for this category include Beijing Opera\cite{b1}, Mridangam Stroke\cite{b2}, Mridangam Tonic\cite{b3}, Nsynth Instrument\cite{b4}, and Nsynth Source \cite{b5}.

    \item \textbf{Acoustic Scene Understanding} This category refers to common urban sounds, such as those produced by buses, cars, jackhammers, drills, dog barks, elevators, crowded streets, subways, trucks, police sirens, motorcycles, air conditioners, engines idling, car horns, and street music as well as sounds commonly encountered in daily life, such as church bells, birds chirping, mouse clicks, ambient office noise, cafes, trams, beaches, restaurants, hens, roosters, metro stations, parks, and city centers. Datasets used for this category include Cochlscene\cite{b7}, ESC50\cite{b8}, TUT\cite{b9} and USD-8K (Urban Sounds) \cite{b10}.

    \item \textbf{Music Genre Classification:} This category involves classifying audio samples into distinct music genres, including classical, country, disco, hip-hop, blues, etc. \cite{b6}.

    \item \textbf{Impact and Emergency Sound:} This category focuses on loud, sudden sounds such as explosions, gunshots, and sirens. The SESA dataset is commonly used for this task \cite{b11}.

    \item \textbf{Non-Verbal Vocalization Sounds:} This category includes non-verbal vocal sounds such as coughing, sneezing, throat clearing, sniffing, sighing, and laughter. Relevant datasets for this category include VocalSound \cite{b12}.
\end{itemize}

\subsection{Task-Specific Prompt Generation}
We begin by providing GPT-4 with information about task categories and their labels, along with examples of sound attributes like `quiet', `loud', `muted', and `faint'. We also include examples of sound sources such as `theater', `concert', `room', `opera', and `street.' We then request GPT-4~\cite{b13} to generate a list of 60 sound attributes and sources relevant to our task categories. Next, we manually map these sound attributes and sources to each task category, ensuring that the attributes and sources are contextually appropriate for the specific labels within each category.

\begin{table*}[t]
\caption{Zero-shot Audio Classification performance of TSPE on five tasks across twelve datasets. Best results are reported in \textbf{bold}.}
\resizebox{\textwidth}{!}{%
\begin{tabular}{llcccc}
\hline \hline
Task & Dataset & MSCLAP 2023 & \begin{tabular}[c]{@{}c@{}}MSCLAP 2023\\ (TSPE)\end{tabular} & MSCLAP 2022 & \begin{tabular}[c]{@{}c@{}}MSCLAP 2022\\ (TSPE)\end{tabular} \\ \hline
\multirow{6}{*}{Musical Instruments Recognition} & Beijing Opera & 70.33 & 68.22 & 55.50 & \textbf{71.86} \\
 & Mridangam Stroke & 45.60 & \textbf{51.77} & 14.69 & 10.72 \\
 & Mridangam Tonic & 20.13 & \textbf{34.01} & 16.52 & 16.20 \\
 & NSynth Instrument & \textbf{66.72} & 64.70 & 27.53 & 30.40 \\
 & NSynth Source & 52.24 & \textbf{53.47} & 38.55 & 34.91 \\ \hline
\multirow{5}{*}{Acoustic Scene Understanding} & Cochlscene & \textbf{85.07} & 83.98 & 22.77 & 24.76 \\
 & USD-8K & 79.37 & \textbf{83.72} & 72.39 & 70.93 \\
 & ESC-50 & 92.85 & \textbf{94.55} & 77.70 & 75.85 \\
 & TUT & 44.07 & \textbf{46.51} & 21.32 & 24.09 \\ \hline
Music Genre Classification & GTZAN & 54.49 & \textbf{59.57} & 22.72 & 19.51 \\ \hline
Impact and Emergency Sound & SESA & 65.71 & 65.71 & 66.28 & \textbf{67.81} \\ \hline
Non-Verbal Vocalization Sound & Vocal Sound (VS) & \textbf{80.93} & 78.94 & 47.09 & 61.23 \\ \hline \hline
\end{tabular}%
}
\label{tab:results}
\end{table*}

For each task category, we then supply GPT-4~\cite{b13} with a prompt format and the list of attributes and sources we have mapped to that category. We ask it to generate 40 prompts using the following formats:
\begin{itemize}
    \item ``A $<$attribute$>$ sound of a $<$label$>$''
    \item ``A sound of a $<$label$>$ coming from a $<$source$>$''
    \item ``A $<$attribute$>$ sound of a $<$label$>$ can be heard from a $<$source$>$''
\end{itemize}
where $<$label$>$ refers to the class label, $<$attribute$>$ refers to the sound attribute, and $<$source$>$ refers to the sound source. From the 40 prompts generated for each task category, we manually filter 20 that best match the task category requirements. We observed some hallucinated or nonspecific prompts from GPT-4~\cite{b13}, making manual selection essential.

\subsection{Hard Prompting and Prompt Ensemble}
As discussed, we employ hard prompting~\cite{b14} over techniques like soft prompting or other learning-based methods because those require retraining or fine-tuning~\cite{b15}, whereas our approach is training-free and enables zero-shot improvement. Our diverse set of prompts, improve downstream performance by describing class labels in various ways, incorporating both their acoustic characteristics and sound sources.

Our method, Task-Specific Prompt Ensemble (TSPE), significantly improves the zero-shot performance of Audio-Language Models for audio classification~\cite{b23}. By using a unique set of prompts for each task category, we capture the subtle nuances and properties of sounds across diverse scenarios. This approach allows for more accurate semantic representation, as prompts like ‘A melodious sound of $<$piano$>$' capture the nuances of the category label better than generic prompts like ‘This is the sound of a $<$label$>$'. For each task category, we generate text embeddings for all prompts in its prompt set and then average these embeddings to enhance semantic representation. In summary, we first create task-specific prompts using hard prompting and then ensemble these prompts to improve zero-shot results.

\subsection{Injecting sound attribute in hard prompting}
From the initial set of 40 prompts generated by GPT-4 for each task category, we manually filter 20 prompts for each category, carefully checking for mismatches in sound attributes. We ensure that the sound attributes in each prompt are meaningfully related to the category labels for the task. For example, the prompt ``A \emph{melodious} sound of a $<$label$>$'' is well-suited for the Music Instruments Classification task, as in ``A \emph{melodious} sound of a $<$piano$>$'', which is semantically appropriate. However, ``A \emph{melodious} sound of a $<$gunshot$>$'' would be mismatched and inappropriate for the Impact and Emergency Sound.

Similarly, prompts like ``A \emph{gentle} sound of an \emph{explosion}'' do not fit the Impact and Emergency Sound and would be better replaced with ``A \emph{gentle} sound of a \emph{flute}'' for the Music Instruments Classification task. This manual filtering step ensures that sound attributes align with the realistic qualities of each task category.

\subsection{Injecting sound source information in hard prompting}
During the manual filtering of the 40 prompts generated by GPT for each task category, we carefully ensure that each sound source aligns naturally with the category labels. This alignment is essential because a mismatched sound source can degrade the coherence of the prompt, reducing its effectiveness in capturing the correct semantic context~\cite{b22}. 

For instance, consider the prompt ``The sound of a \emph{violin} can be heard from an \emph{orchestra}.'' This is a more plausible pairing than alternatives like ``The sound of a \emph{violin} coming from a \emph{library}'' or `a \emph{zoo}', where the source does not fit the expected environment for a violin. Similarly, the prompt ``The sound of an \emph{organ} coming from the \emph{church}'' aligns well with the category label `\emph{organ}', providing a realistic setting. In contrast, options like ``The sound of an organ coming from an \emph{airport}'' or a ``\emph{railway station}'' would seem out of place, reducing the prompt's effectiveness in discriminating organ sound from a pool of sounds.

\section{Experimental Setup}
We evaluate our method across five task categories using twelve different datasets and two state-of-the-art Audio Language Models (ALMs), MS-CLAP'22 and MS-CLAP'23 \cite{b19}, both developed by Microsoft and released in 2022 and 2023, respectively. For the Musical Instrument Recognition task, we test our model on multiple audio datasets that include a wide range of musical instruments. These datasets include Beijing Opera \cite{b1}, Mridangam Stroke \cite{b2}, Mridangam Tonic \cite{b3}, Nsynth Instrument \cite{b4}, and Nsynth Source \cite{b5}. For the Music Genre Classification task, we use the GTZAN dataset \cite{b6}, which contains music from various genres, including Rock, Hip-Hop, Pop, and Country. For Acoustic Scene Understanding, we evaluate our model on multiple datasets, such as Cochlscene \cite{b7}, USD-8K (Urban Sounds) \cite{b10}, ESC50 \cite{b8}, and TUT Urban Acoustic Scenes \cite{b9}. These datasets contain diverse sounds like air conditioners, car horns, trams, and metro stations. For Impact and Emergency Sound Classification, we use the SESA dataset \cite{b11}, which includes emergency-related sounds like explosions, gunshots, and sirens. For Non-Verbal Vocalization Sound Classification, we test on the VocalSound dataset \cite{b12}, which includes sounds such as sighing, sniffing, laughing, and sneezing. All results are reported by taking average across five runs.
\section{Results}
Table~\ref{tab:results} presents the effectiveness of our technique for Zero-Shot Audio Classification (ZSAC) applied to two state-of-the-art Audio-Language Models (ALMs), MSCLAP'22 and MSCLAP'23\cite{b19}. We provide a comparison of our Task-Specific Prompt Ensemble method against vanilla prompts for ZSAC across five task categories and twelve audio classification datasets covering diverse sounds.

\subsection{Result Analysis}
TSPE is able to improve the performance of the models ranging from 1.23\% to 16.36\%, with an average improvement of 2.06\% across all datasets on MSCLAP'23 and 1.89\% on MSCLAP'22. We observe that on some datasets, performance decreases after applying TSPE. One reason could be that the prompts are not linguistically rich enough for that task and we can take this up as future work. Table~\ref{tab:prompt_example} shows examples of prompts generated by TSPE for different tasks.

\subsection{Hyper-Parameter Tuning}
\begin{figure}[h!]
    \centering
    \includegraphics[width=0.5\textwidth]{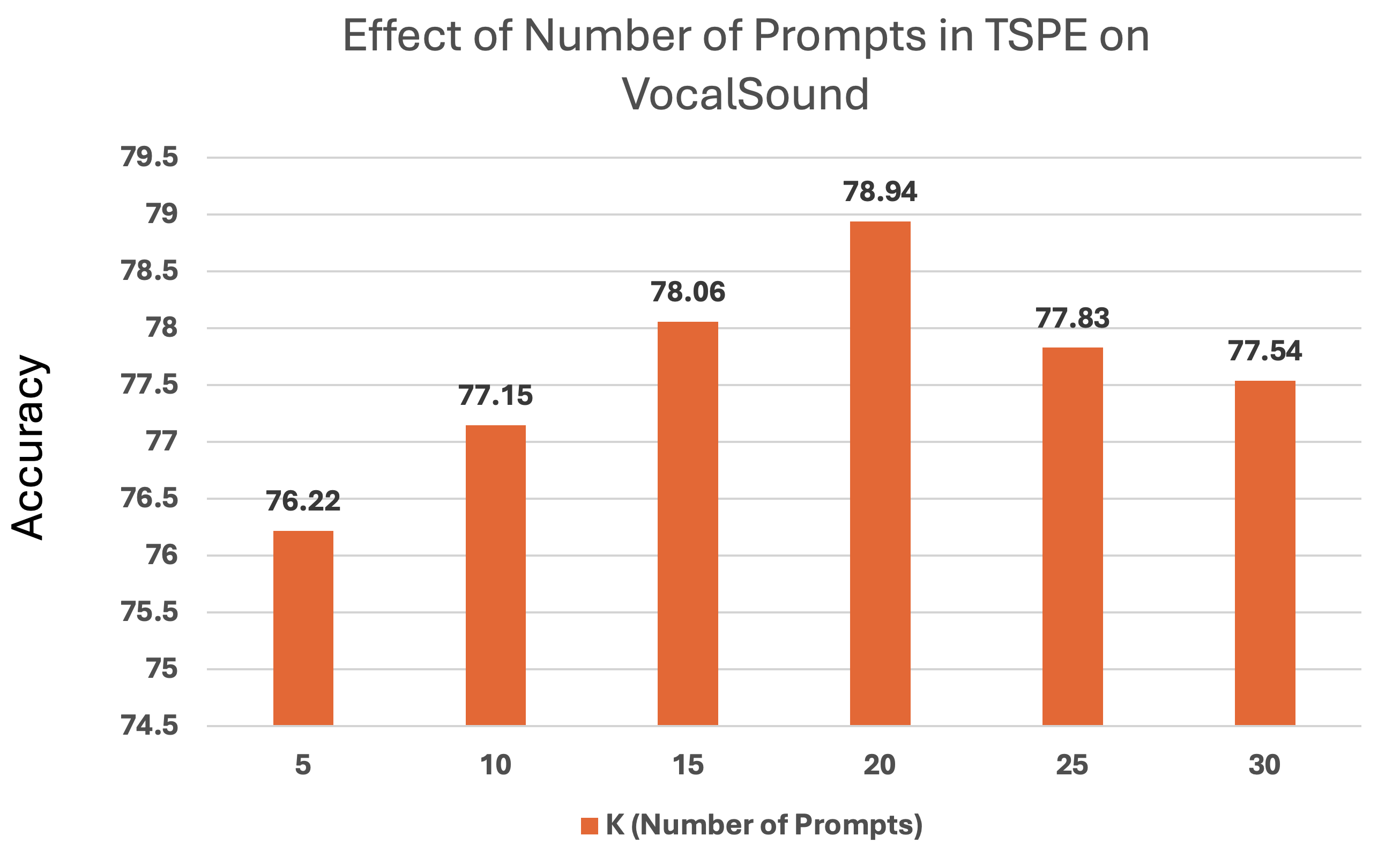}
    \caption{Effect of different number of prompts on MSCLAP'23 for Audio Classification on the VocalSound dataset}
    \label{fig:ablation}
\end{figure}

We select \( K = 20 \) prompts from an initial set of 40 generated by GPT-4, and perform an ablation study to determine the optimal value for \( K \). Specifically, we evaluate the performance of TSPE on the VocalSound dataset with \( K \) values of \{5, 10, 15, 20, 25, 30\} to examine how the number of prompts affects the performance of Audio-Language Models (ALMs) on zero-shot audio classification. Results indicate that ALM performance improves as \( K \) increases up to 20, after which it declines. This drop in performance beyond \( K = 20 \) may result from increased semantic noise due to the higher number of prompts. Fig.~\ref{fig:ablation} illustrates the effect of different prompt counts on MSCLAP'23 for audio classification on the VocalSound dataset \cite{b12}.

\begin{table}[]
\caption{Task Categories and prompt examples for them}
\resizebox{0.48\textwidth}{!}{%
\begin{tabular}{|l|l|}
\hline
Task                                             & Prompt Example                                                              \\ \hline
\multirow{2}{*}{Musical Instruments Recognition} & The sound of a $<$violin$>$ coming from an $<$opera$>$ \\
                                                 & The sound of an $<$organ$>$ coming from a $<$church$>$            \\ \hline
\multirow{2}{*}{Acoustic Scene Understanding}    & A $<$loud$>$ sound of a $<$jackhammer$>$ coming from a $<$street$>${}      \\
                                                 & The sound of a $<$bike$>$ coming from a $<$road$>$                 \\ \hline
\multirow{2}{*}{Music Genre Classification}      & The sound of $<$jazz$>$ coming from a $<$concert hall$>$       \\
                                                 & The sound of $<$rock$>$ coming from a $<$room$>$         \\ \hline
\multirow{2}{*}{Impact and Emergency Sound}      & The sound of $<$gunshot$>$ coming from a $<$university$>$  \\
                                                 & A sound of an $<$explosion$>$ coming from a $<$parking lot$>$    \\ \hline
\multirow{2}{*}{Non-Verbal Vocalization Sound}   & A $<$hushed$>$ sound of a $<$cough$>${}                          \\
                                                 & A sound of $<$laughter$>$ coming from a $<$hall$>$             \\ \hline
\end{tabular}%
}
\label{tab:prompt_example}
\end{table}

\section{Conclusion}
In this paper, we present TSPE, a task-specific hard prompting method designed to enhance the zero-shot performance of existing state-of-the-art audio language models (ALMs). Unlike standard prompts commonly used in the literature, our approach creates task-specific prompts by first analyzing the individual class labels across various audio classification tasks. We then identify the most relevant sound attributes and sources that naturally define each class label in a natural language. Finally, we conduct extensive quantitative and qualitative experiments, demonstrating that TSPE significantly outperforms traditional zero-shot evaluation.

\section{Limitation and Future Work}
\begin{enumerate}
    \item Prompt Generation Errors: Errors or repetitive phrasing in GPT-4-generated prompts may require careful manual filtering. Future work will explore automated quality control methods to reduce the need for human oversight.
    
    \item Bias in Task-Specific Prompts: The customization of prompts may introduce task-specific biases into the model. Future efforts will focus on identifying and mitigating these biases to ensure robust performance across diverse datasets.
    
    \item Broader Application of TSPE Representations: TSPE's text-audio representations could enhance other tasks, such as audio generation and sound event localization. Future research will include extending TSPE to these applications.
\end{enumerate}



\begin{thebibliography}{00}
\bibitem{b1} Tian, Mi, et al. "A study of instrument-wise onset detection in Beijing opera percussion ensembles." 2014 ieee international conference on acoustics, speech and signal processing (icassp). IEEE, 2014.
\bibitem{b2}  Akshay Anantapadmanabhan, Ashwin Bellur, and Hema A. Murthy. 2014. Mridangam stroke dataset (1.0). 2013 IEEE International Conference on Acoustics, Speech and Signal Processing, Vancouver, BC, Canada. Data set.
\bibitem{b3} Akshay Anantapadmanabhan, Ashwin Bellur, and Hema A. Murthy. 2014. Mridangam stroke dataset (1.0). 2013 IEEE International Conference on Acoustics, Speech and Signal Processing, Vancouver, BC, Canada. Data set.
\bibitem{b4} Engel, Jesse, et al. "Neural audio synthesis of musical notes with wavenet autoencoders." International Conference on Machine Learning. PMLR, 2017.
\bibitem{b5} Engel, Jesse, et al. "Neural audio synthesis of musical notes with wavenet autoencoders." International Conference on Machine Learning. PMLR, 2017.
\bibitem{b6} Tzanetakis, George, and Perry Cook. "Musical genre classification of audio signals." IEEE Transactions on speech and audio processing 10.5 (2002): 293-302.
\bibitem{b7} Jeong, Il-Young, and Jeongsoo Park. "CochlScene: Acquisition of acoustic scene data using crowdsourcing." 2022 Asia-Pacific Signal and Information Processing Association Annual Summit and Conference (APSIPA ASC). IEEE, 2022.
\bibitem{b8} Karol J. Piczak. ESC: Dataset for Environmental Sound Classification. In Proceedings of the 23rd Annual ACM Conference on Multimedia, pages 1015–1018. ACM Press.
\bibitem{b9} Annamaria Mesaros, Toni Heittola, and Tuomas Virtanen. 2017b. Tut acoustic scenes 2017, development dataset. Data set.
\bibitem{b10} Salamon, Justin, Christopher Jacoby, and Juan Pablo Bello. "A dataset and taxonomy for urban sound research." Proceedings of the 22nd ACM international conference on Multimedia. 2014.
\bibitem{b11} Tito Spadini. 2019. Sound events for surveillance applications (1.0.0). Data set.
\bibitem{b12} Gong, Yuan, Jin Yu, and James Glass. "Vocalsound: A dataset for improving human vocal sounds recognition." ICASSP 2022-2022 IEEE International Conference on Acoustics, Speech and Signal Processing (ICASSP). IEEE, 2022.
\bibitem{b13} Achiam, Josh, et al. "Gpt-4 technical report." arXiv preprint arXiv:2303.08774 (2023).
\bibitem{b14} Jin, Woojeong, et al. "A good prompt is worth millions of parameters: Low-resource prompt-based learning for vision-language models." arXiv preprint arXiv:2110.08484 (2021).
\bibitem{b15} Zhou, Kaiyang, et al. "Learning to prompt for vision-language models." International Journal of Computer Vision 130.9 (2022): 2337-2348.
\bibitem{b16}Radford, Alec, et al. "Learning transferable visual models from natural language supervision." International conference on machine learning. PMLR, 2021.
\bibitem{b17} Wu, Ho-Hsiang, et al. "Wav2clip: Learning robust audio representations from clip." ICASSP 2022-2022 IEEE International Conference on Acoustics, Speech and Signal Processing (ICASSP). IEEE, 2022.
\bibitem{b18} Guzhov, Andrey, et al. "Audioclip: Extending clip to image, text and audio." ICASSP 2022-2022 IEEE International Conference on Acoustics, Speech and Signal Processing (ICASSP). IEEE, 2022.
\bibitem{b19} Elizalde, Benjamin, et al. "Clap learning audio concepts from natural language supervision." ICASSP 2023-2023 IEEE International Conference on Acoustics, Speech and Signal Processing (ICASSP). IEEE, 2023.
\bibitem{b20} Sahoo, Pranab, et al. "A systematic survey of prompt engineering in large language models: Techniques and applications." arXiv preprint arXiv:2402.07927 (2024).
\bibitem{b21} Ghosh, Sreyan, et al. "Compa: Addressing the gap in compositional reasoning in audio-language models." arXiv preprint arXiv:2310.08753 (2023).
\bibitem{b22} Khattak, Muhammad Uzair, et al. "Maple: Multi-modal prompt learning." Proceedings of the IEEE/CVF Conference on Computer Vision and Pattern Recognition. 2023.
\bibitem{b23} Zaman, Khalid, et al. "A survey of audio classification using deep learning." IEEE Access (2023).
\bibitem{b24} Wen, Yuxin, et al. "Hard prompts made easy: Gradient-based discrete optimization for prompt tuning and discovery." Advances in Neural Information Processing Systems 36 (2024).
\bibitem{b25} Li, Yiming, Xiangdong Wang, and Hong Liu. "Audio-Free Prompt Tuning for Language-Audio Models." ICASSP 2024-2024 IEEE International Conference on Acoustics, Speech and Signal Processing (ICASSP). IEEE, 2024.
\bibitem{b26} Liang, Jinhua, et al. "Adapting language-audio models as few-shot audio learners." arXiv preprint arXiv:2305.17719 (2023).
\bibitem{b27} Wu, Jiayang, et al. "Multimodal large language models: A survey." 2023 IEEE International Conference on Big Data (BigData). IEEE, 2023.
\bibitem{b28} Bai, Tianyi, et al. "A Survey of Multimodal Large Language Model from A Data-centric Perspective." arXiv preprint arXiv:2405.16640 (2024).
\bibitem{b29} Jin, Yizhang, et al. "Efficient multimodal large language models: A survey." arXiv preprint arXiv:2405.10739 (2024).
\bibitem{b30} Szot, Andrew, et al. "Grounding Multimodal Large Language Models in Actions." arXiv preprint arXiv:2406.07904 (2024).
\bibitem{b31} Carolan, Kilian, Laura Fennelly, and Alan F. Smeaton. "A Review of Multi-Modal Large Language and Vision Models." arXiv preprint arXiv:2404.01322 (2024).

\end{thebibliography}
\end{document}